\documentclass[a4paper]{article}
\input epsf
\begin {document}

\title {\bf Nodes in the Relativistic Quantum Trajectories and
Photon's Trajectories}
\author{ T.~Djama\footnote{Electronic address:
{\tt djam\_touf@yahoo.fr}}}

\date{November 4, 2003}

\maketitle

\begin{center}Universit\'e des Sciences et de la
Technologie
Houari Boum\'edienne, \\Alger, Algeria \\
\vspace*{0.5cm}
Mail address:14, rue Si El Hou\`es,
B{\'e}ja{\"\i}a, Algeria
\end{center}

\begin{abstract}
\noindent
Through the constant potential and the linear potential, we
establish the existence of nodes for the relativistic quantum
trajectories as the same way as for the quantum trajectories.
We establish the purely relativistic limit $(\hbar \to 0)$
for these trajectories, and link the nodes to de Broglie's
wavelength.
\end{abstract}

\vskip\baselineskip



\newpage

% %%%%%%%%%%%%%%%%%%%%%%%%%%%%%%%%%%%%%%%%%%%%%%%%%%%%%%%%%%%%%%%%%
\vskip0.5\baselineskip
\noindent
{\bf 1.\ \ Introduction }
\vskip0.5\baselineskip
% %%%%%%%%%%%%%%%%%%%%%%%%%%%%%%%%%%%%%%%%%%%%%%%%%%%%%%%%%%%%%%%%%

Recently, we have presented with Bouda a new formulation of
quantum mechanics based on both new quantum Lagrangian and the
solution of the quantum stationary Hamilton-Jacobi equation
(QSHJE) \cite{Dja1}. The Bouda and Djama Lagrangian lead to the
quantum Newton's law, from which they derived and plotted the
quantum trajectories (QT) for several potentials \cite{Dja2}. In
a previous paper \cite{Dja3}, we have generalized the formulation
presented in Ref.  \cite{Dja1,Dja2} into the relativistic quantum
systems. We started from the relativistic quantum stationary
Hamilton-Jacobi equation (RQSHJE)
presented by Faraggi and Matone in Refs. \cite{FM1,FM2,FM3} as%
\begin{eqnarray}
{1 \over 2m_0}\left({\partial S_0 \over \partial
x}\right)^2-
{\hbar^2 \over 4m_0}\left[{3 \over 2}
\left({\partial S_0 \over \partial x}\right)^{-2}
\left({\partial^2 S_0 \over \partial
x^2}\right)^2-\right.
\hskip35mm&& \nonumber\\
\left.\left({\partial S_0 \over \partial
x}\right)^{-1}
\left({\partial^3 S_0 \over \partial x^3}\right)
\right]+
{1 \over 2m_0c^2}\left[m_0^2c^4
-(E-V)^2\right]=0\; ,
\end{eqnarray}
where $S_0$ represent the relativistic quantum reduced action,
$E$ the total energy including the energy at rest, $V$ the
potential and $m_0$ the mass at rest of the particle. $c$ is the
light velocity in vacuum. The relativistic quantum reduced action
is given by  \cite{Dja1,FM1,FM2,FM3}
\begin {equation}
S_0=\hbar \ \arctan {\left [ a  {\phi_1
\over \phi_2 } +b \right ]}  \; .
\end {equation}
$a$ and $b$ are real integration constants. $\phi_1$ and
$\phi_2$ are two real independents solutions of the
Klein-Gordon equation
\begin{equation}
-c^2 \hbar^2 {\partial^2 \phi \over \partial x^2}+
\left[m_0^2c^4-(E-V)^2\right]\phi(x)=0\; .
\end{equation}
The conjugate momentum can be written following Eq. (2) as
\begin{equation}
{\partial S_0 \over \partial x}=
\pm {\hbar a W \over \phi_2^2+(a\phi_1+b\phi_2)^2} \; .
\end{equation}
$W$ represent the wronskian of $\phi_1$ and $\phi_2$
with respect to $x$. The sign $\pm$ in Eq. (4) represent
the two  possibilities of motion: $x$ positive or $x$ negative
directions.

Taking advantage on this results, we introduced in Ref.
\cite{Dja3} a relativistic  quantum Lagrangian
\begin{equation}
L=-m_0c^2 \sqrt{1-{\dot{x}^2 \over c^2}\;
f(x,\mu,\nu,E)}-V(x)\; ,
\end{equation}
from which we derived the first integral of the relativistic
quantum Newton's Law (FIRQNL).
\begin{eqnarray}
\left[(E-V)^2-m_0^2c^4\right]^2+{\dot{x}^2 \over
c^2}(E-V)^2
\left[(E-V)^2-m_0^2c^4\right]+{\hbar^2 \over 2}
\left[{3 \over 2}\left({\ddot{x} \over
\dot{x}}\right)^2-
{\dot{\ddot{x}} \over \dot{x}}\right] \cdot
\hskip-10mm&& \nonumber\\
(E-V)^2-{\hbar^2 \over 2}\left(\ddot{x}{dV \over dx}+
\dot{x}^2{d^2V \over dx^2}\right)
\left[{(E-V)^2+m_0^2c^4 \over (E-V)^2-m_0^2c^4}\right]
(E-V)^2-
{3\hbar^2 \over 4}\cdot
\hskip-1mm&& \nonumber\\
\left(\dot{x}{dV \over dx}\right)^2
\left[{(E-V)^2+m_0^2c^4 \over
(E-V)^2-m_0^2c^4}\right]^2-
\hbar^2\left(\dot{x}{dV \over dx}\right)^2{m_0^2c^4
\over (E-V)^2-m_0^2c^4}
=0\; .
\end{eqnarray}
We also established that the conjugate momentum of relativistic
quantum systems is connected with the particle velocity
as follows
\begin{equation}
{\partial S_0 \over \partial x}={E-V(x) \over
\dot{x}}-
{m_0^2 c^4 \over (E-V)\dot{x}} ,
\end{equation}
From Eqs. (4) and (7), one can write the following
expressions
\begin{equation}
{dx \over dt}=\pm {1 \over \hbar a W}\left[E-V(x)-{m_0^2c^4 \over
E-V(x)} \right] \left[\phi_2^2+(a\phi_1+b\phi_2)^2\right] \; .
\end{equation}
We choose the sign of $aW$ in such a way that in classically
allowed region $(E-V>m_0c^2)$, in Eq. (8) $dx/dt$ takes the
positive sign for $x$ positive directions and the negative
sign for the $x$ negative directions. Furthermore, at the
turning point $dx/dt$ changes the sign either the particle
goes in classically allowed regions or in the classically
forbidden regions.

% %%%%%%%%%%%%%%%%%%%%%%%%%%%%%%%%%%%%%%%%%%%%%%%%%%%%%%%%%%%%%%%%%
\vskip0.5\baselineskip
\noindent
{\bf 2.\ \ Motion of electron under the constant potential }
\vskip0.5\baselineskip
% %%%%%%%%%%%%%%%%%%%%%%%%%%%%%%%%%%%%%%%%%%%%%%%%%%%%%%%%%%%%%%%%%

In the case of an electron moving under a constant potential
$V=U_0$ , the FIRQNL (Eq. (6)) and the conjugate momentum
(Eq. (7)) reduce to the following equations
\begin{eqnarray}
\left[(E-U_0)^2-m_0^2c^4\right]^2-{\dot{x}^2 \over c^2}\;
(E-U_0)^2 \; \left[(E-U_0)^2-\right.
\hskip25mm\nonumber\\
\left. m_0^2c^4\right]+ {\hbar^2 \over 2}\left[{3 \over
2}\left({\ddot{x} \over \dot{x}}\right)^2- {\dot{\ddot{x}} \over
\dot{x}}\right](E-U_0)^2=0\; .
\end{eqnarray}
and
\begin{equation}
\dot{x}{\partial S_0 \over \partial x}=E-U_0
-{m_0^2c^4 \over E-U_0}\; .
\end{equation}
The solution of these equation is
\begin {equation}
x(t)={\hbar c \over \sqrt{E^2-m_0^2c^4}}
\arctan{\left[a \tan{\left({E^2-m_0^2c^4
\over\hbar E}\; t\right)}+b\right]}+{\pi \hbar c \over
\sqrt{E^2-m_0^2c^4}}n+x_0  \; .
\end {equation}
with
$$
t\; \in \left[{\pi \hbar E  \over
E^2-m_0^2c^4}\left(n-{1 \over 2}\right)
; {\pi \hbar E  \over E^2-m_0^2c^4}\left(n+{1 \over
2}\right) \right]
$$
%
%
%%%%%%%%%%%%%%%%%%%%%%%%%%%%%%%%%%%%%%%%%%%%%%%%%%%%%%%%%%%%%%%%%%%
\begin{center}
\begin{figure}
\def\put(#1,#2)#3{\leavevmode\rlap{\hskip#1\unitlength\raise#2\unitlength\hbox{#3}}}
\centerline{ \vbox{\hsize=10.5cm\setlength{\unitlength}{1truecm}
\put(0,0){\epsfxsize=10cm \epsfbox{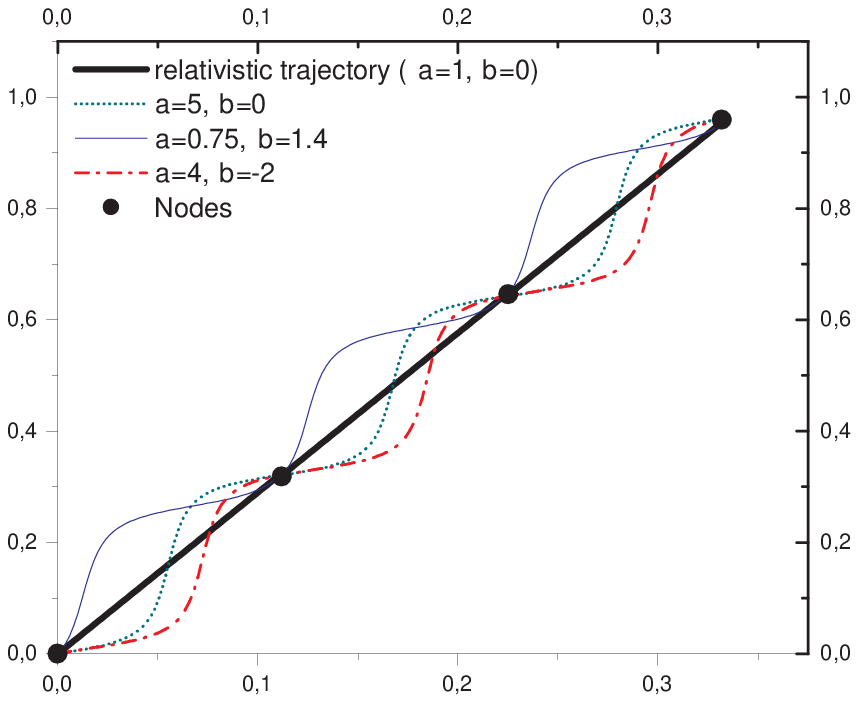}}
\put(8.5,0.6){$t$($\times 10^{-20}$ s)}
\put(0.2,7.75){$x$($\times 10^{-12}$ m)}}}
\centerline { \vbox {
\hsize=12cm\noindent \small Fig. 1:  Relativistic quantum
trajectories for a free electron of energy $E=2$ Mev. For all the
curves, we have chosen $x(t=0)= 0$.} }
\end{figure}
\end{center}
for every integer number. In Eq. (11) we added the term multiple
of $n$ to assure that the particle will not contained between
$$
-{\hbar c  \over \sqrt{(E-U_0)^2-m_0^2c^4}} {\pi \over 2}+x_0
$$
and
$$
{\hbar c  \over \sqrt{(E-U_0)^2-m_0^2c^4}} {\pi \over 2}+x_0
\; .
$$
Note that for $a=1$ and $b=0$ the relation (11) reduces to
the relativistic relation
\begin {equation}
x(t)= {c \over E-U_0}\; \sqrt{(E-U_0)^2-m_0^2c^4}\; t+x_0  \; .
\end {equation}
In Fig. 1, we have plotted in $(t,x)$ plane for an electron of
energy $2Mev+U_0$ some trajectories for different values of
$a$ and $b$. All these trajectories pass through nodes
exactly as we have seen for quantum trajectories of an electron
moving under a constant potential \cite{Dja3}.
These node correspond to the times
\begin {equation}
t_n={\pi \hbar E-U_0  \over (E-U_0)^2-m_0^2c^4}\left(n+{1 \over
2}\right)\; .
\end {equation}
The distance between two adjacent nodes is on time
axis
\begin {equation}
\Delta t_n=t_{n+1}-t_n={\pi \hbar (E-U_0)  \over
(E-U_0)^2-m_0^2c^4}\; .
\end {equation}
and space axis
\begin {equation}
\Delta x_n=x_{n+1}-x_n={\pi \hbar c  \over
\sqrt{(E-U_0)^2-m_0^2c^4}}\; .
\end {equation}
These distances are both proportional to $\hbar$
meaning that at
the classical limit $\hbar \to 0$ the nodes become
infinitely
near, and then, all possible relativistic quantum trajectories
tend to the purely relativistic one \cite{Dja1,Dja2}.
Indeed, let us consider   an arbitrary point $P(x,t)$ from
any relativistic trajectory. Obviously, this point is situated
between two nodes. Now, if we take the orthogonal projection
$P_0$ of $P$ on the relativistic trajectory ($a=1$, $b=0$) and
compute the distance $PP_0$ we get
\begin{equation}
PP_0=c\;
\sqrt{2-{(E-U_0)^2-m_0^2c^4 \over (E-U_0)^2}} \|t_P-t_{P_0}\|\; .
\end {equation}
Since Eq. (8) indicates that $\dot{x}$ is a monotonous
function, $ \|t_P-t_{P_0}\|<t_{n+1}-t_{n}$ for every
relativistic quantum trajectory \cite{Dja2}. Then at the
classical limit $(\hbar \to 0)$, $\|t_P-t_{P_0}\| \to 0$ and
$PP_0 \to 0$. Thus, all quantum trajectories goes when
$(\hbar \to 0)$ to the purely relativistic one. This result is
with accord with the fact that our dynamical equations
(Eqs (1), (6) and (7)) tend to the relativistic equation
when $(\hbar \to 0)$.

Now let us consider the problem when $E-U_0<0$
when the electron is in the classically forbidden regions.
For this case the solution of Eqs. (9) and (10) is
\begin{equation}
x(t) = {\hbar \; c \over 2\sqrt{m_0^2c^4-(E-U_0)^2}}\ln
{\left\vert
a \tan \left({m_0^2c^4-(E-U_0)^2
\over\hbar E-U_0}\; t\right)+b\right\vert}
+ x_0 \; .
\end{equation}
where $a$, $b$ and $c$ are real integration constant
with $a\neq 0$. We see clearly from Eq. (17) that at the
finite time $-(2n+1)\pi \hbar E /4((E-U_0)^2-m_0^2c^4)$
the electron cross an infinite distance and
reach an infinite speed. This is in accordance with standard
quantum tunneling theories which predict infinite
velocities and finite reflexion times for tunneling
phenomena (Fletcher \cite{fle}, Hartman \cite{hart}).

In Fig. (2), we plotted  for an electron of energy
$2Mev-U_0$ a relativistic quantum trajectory in the
classically forbidden regions with
$a=4$ and $b=2$. This figure shows clearly how the
particle reach an infinite position at a finite time.
In particular, as it is shown by Eq. (17), classically
forbidden regions there is no nodes for the RQT.

%%%%%%%%%%%%%%%%%%%%%%%%%%%%%%%%%%%%%%%%%%%%%%%%%%%%%%%%%%%%%%%%%%%

\begin{figure}
\def\put(#1,#2)#3{\leavevmode\rlap{\hskip#1\unitlength\raise#2\unitlength\hbox{#3}}}
\centerline{ \vbox{\hsize=10.5cm \setlength{\unitlength}{1truecm}
\put(0,0){\epsfxsize=10cm \epsfbox{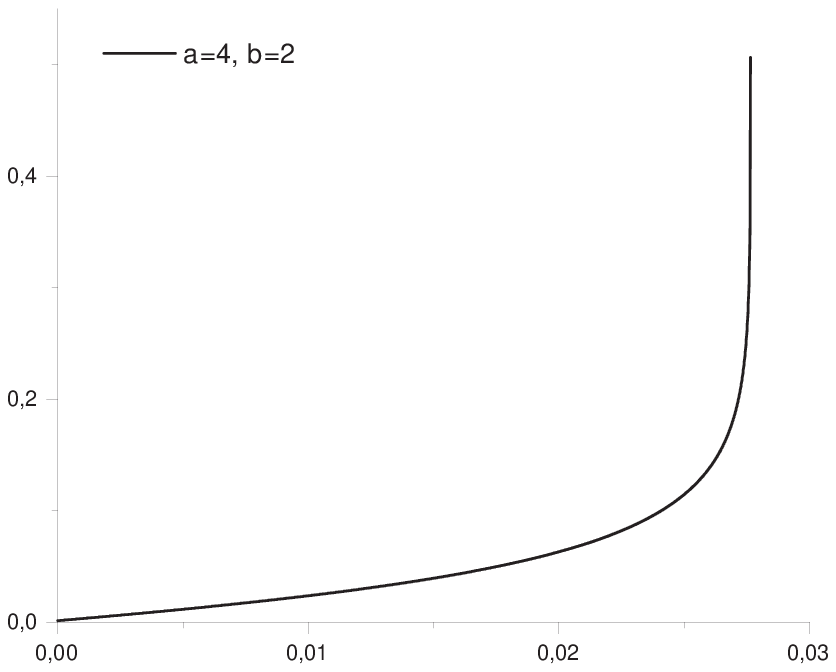}}%
\put(8.5,0.6) {$t$ ($\times 10^{-20}$ s)}%
\put(0.2,7.75) {$x$ ($\times 10^{-12}$ m)}%
}}
%\medskip
%\scriptstyle
\centerline { \vbox
{ \hsize=12cm\noindent
\small Fig. 2:  Relativistic quantum trajectories for
an electron of energy $E=2-U_0$ Mev moving in the
classically forbidden regions. We have chosen
$x(t=0)= 0$.}
}
\end{figure}

\newpage

% %%%%%%%%%%%%%%%%%%%%%%%%%%%%%%%%%%%%%%%%%%%%%%%%%%%%%%%%%%%%%%%%%
\vskip0.5\baselineskip
\noindent
{\bf 3.\ \ Photon trajectories }
\vskip0.5\baselineskip
% %%%%%%%%%%%%%%%%%%%%%%%%%%%%%%%%%%%%%%%%%%%%%%%%%%%%%%%%%%%%%%%%%

We propose, now, to study the motion of a photon
in a presence of a constant potential $U_0$. For this aim,
we must take into consideration that the photon's mass
at rest is null, otherwise its relativistic mass were be
infinite. Then, Eqs. (6) and (7) can be written as

\begin {equation}
(E-U_0)^2-\dot{x}^2{(E-U_0)^2 \over c^2}+
{\hbar^2 \over 2}
\left[{3 \over 2}\left({\ddot{x} \over
\dot{x}}\right)^2-
{\dot{\ddot{x}} \over \dot{x}}\right]=0\; .
\end {equation}
and
\begin {equation}
\dot{x} {\partial S_0 \over \partial x}= E-U_0\; .
\end {equation}
Before introducing the solution of Eq. (18), note that Eq. (19)
indicates that $\dot{x}$ and $\partial S_0 / \partial x$ have the
same sign in classically permitted regions, and are opposite in
classically forbidden regions. Then,  Eq. (19) reduces, after
using the solutions $\cos \left({\|E-U_0\| \over \hbar c}x
\right)$ and $\sin \left({\|E-U_0\|\over \hbar c}x \right)$of Eq.
(3), to
\begin{equation}
{dx \over dt} = \pm {c \over a} \left[ \cos^2 \left({ \|E-U_0\|
\over \hbar c } \right) +
 \left[a \, \sin \left({\|E-U_0\|\over \hbar c}x \right) +
b \, \cos \left({\|E-U_0\|\over \hbar c}x \right)\right]^2
    \right]
\end{equation}
where the sign $+$ is used for the classically permitted regions
and the sign $-$ is used for the forbidden.

As we can deduce from Eq. (20) the motion of a photon under
a constant potential is identical in the two quantum possible
situations, the classically forbidden regions and permitted
regions. It is essentially due to the symmetry that the
Klein-Gordon equation shows for the photon since its solutions
are trigonometric equation either for $E-U_0>0$ or $E-U_0<0$.
The resolution of Eq. (20) gives  \cite{Dja3}
\begin{equation}
x(t)={\hbar c \over \|E-U_0\|} \arctan\left[a
\tan\left({\|E-U_0\| \over \hbar }\; t\right)+ b\right]+{\pi
\hbar c \over \|E-U_0\|}\; n+x_0  \; ,
\end{equation}
with
$$
t \in \left[{\pi \hbar \over \|E-U_0\|}(n-{1 \over 2}),
{\pi \hbar \over \|E-U_0\|}(n+{1 \over 2})\right]
$$

Note that for $a=1$ and $b=0$, Eq. (21) reduces to
\begin {equation}
x(t)=ct+x_0\; ,
\end {equation}
representing the relativistic equation of the photon's
trajectories. As for the massive particle case we note
that the RQTs of the photon pass through some points
constituting nodes. All the trajectories cross the distance
between two nodes at equal times. This establishment is
also valid for the massive particle RQTs.
However, the photon's RQTs in classically forbidden regions
possess also nodes. This is not the case of the massive
particle RQTs in these regions (see the precedent section).

To illustrate, we plotted in Fig. 3 many trajectories, of a free
photon of energy $1.2$ Mev (X ray), for different values
of $(a,b)$. The nodes are clearly
illustrated. The physical interpretation of these nodes
will be exposed in Sec. 5.

%%%%%%%%%%%%%%%%%%%%%%%%%%%%%%%%%%%%%%%%%%%%%%%%%%%%%%%%%%%%%%%%%%%

\begin{figure}
\def\put(#1,#2)#3{\leavevmode\rlap{\hskip#1\unitlength\raise#2\unitlength\hbox{#3}}}
\centerline{ \vbox{\hsize=10.5cm \setlength{\unitlength}{1truecm}
\put(0,0){\epsfxsize=10cm \epsfbox{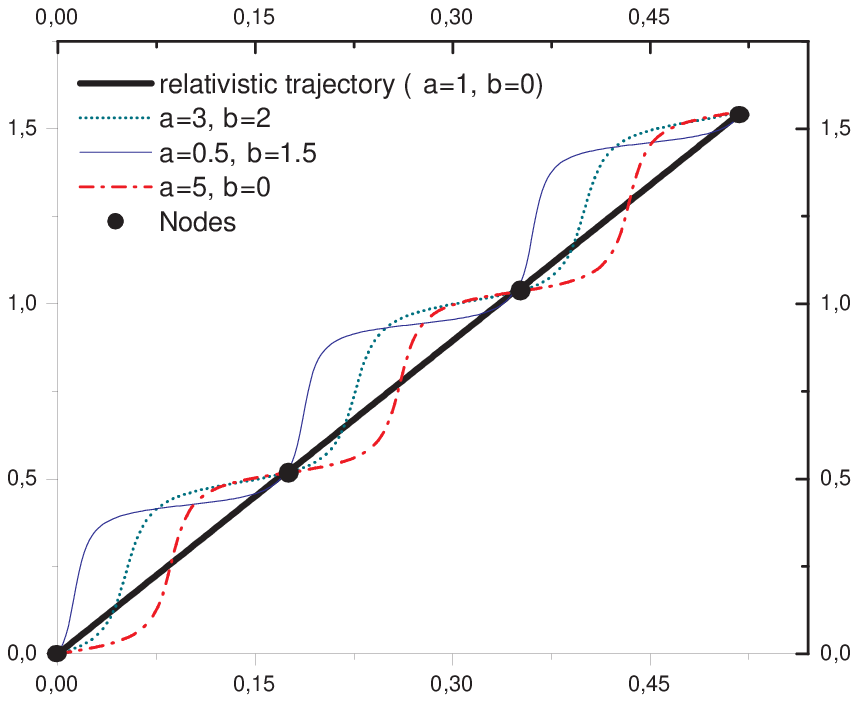}}%
\put(8.5,0.6) {$t$ ($\times 10^{-20}$ s)}%
\put(0.2,7.75) {$x$ ($\times 10^{-12}$ m)}%
}}
%\medskip
%\scriptstyle
\centerline { \vbox { \hsize=12cm\noindent \small Fig. 3:
Relativistic quantum trajectories for a free photon of energy
$E=1.2$ Mev. We have chosen $x(t=0)= 0$.} }
\end{figure}

% %%%%%%%%%%%%%%%%%%%%%%%%%%%%%%%%%%%%%%%%%%%%%%%%%%%%%%%%%%%%%%%%%
\vskip0.5\baselineskip
\noindent
{\bf 4.\ \ The linear potential case }
\vskip0.5\baselineskip
% %%%%%%%%%%%%%%%%%%%%%%%%%%%%%%%%%%%%%%%%%%%%%%%%%%%%%%%%%%%%%%%%%

Here, we investigate the motion of a massive particle
(electron) under a potential of the form
\begin {equation}
V(x) = gx  \; ,
\end {equation}
for which the Klein-Gordon equation takes the form
\begin{equation}
-c^2 \hbar^2 {\partial^2 \phi \over \partial x^2}+
\left[m_0^2c^4-(E-gx)^2\right]\phi(x)=0\; .
\end{equation}
To establish the RQTs for the linear potential case, we can
integrate the Eq. (8), where $\phi_1$ and $\phi_2$ are, now,
two solutions of Eq. (24).

In this paper, we do not present the analytic solutions of
Eq. (24), and in order to plot the RQTs, we approach the
problem by numeric methods. We, first, integrate numerically
Eq. (24) to obtain two independents solutions
$\phi_1$ and $\phi_2$, then, we plot the RQTs from Eq. (23).
We opt in the two steps for the Euler integration method.

\noindent The RQTs for the linear potential are presented in Fig.
4. We choose $E=2$ Mev, and $g=1$ Kg.m.s. In Ref. \cite{Dja2}, we
have chosen $g=10^{-9}$ Kg.m.s, which is very small compared with
$g=1$ Kg.m.s. We take this last value for relativistic problem to
render the quantity $gx$, for quantum scales, of the same order
as $E$, so that the Klein-Gordon equation do not reduces to the
Schr\"odinger equation.

As we can notice from Fig. 4, the nodes are also present
for the linear potential case. The distance between two
adjacent nodes in Fig. 4 increase as the velocity decrease
when it approaches the turning point (point where the
velocity vanish). This note will be exposed in Sec. 5.
Here we do not investigate the classically forbidden regions.
%%%%%%%%%%%%%%%%%%%%%%%%%%%%%%%%%%%%%%%%%%%%%%%%%%%%%%%%%%%%%%%%%%%

\begin{figure}
\def\put(#1,#2)#3{\leavevmode\rlap{\hskip#1\unitlength\raise#2\unitlength\hbox{#3}}}
\centerline{ \vbox{\hsize=10.5cm \setlength{\unitlength}{1truecm}
\put(0,0){\epsfxsize=10cm \epsfbox{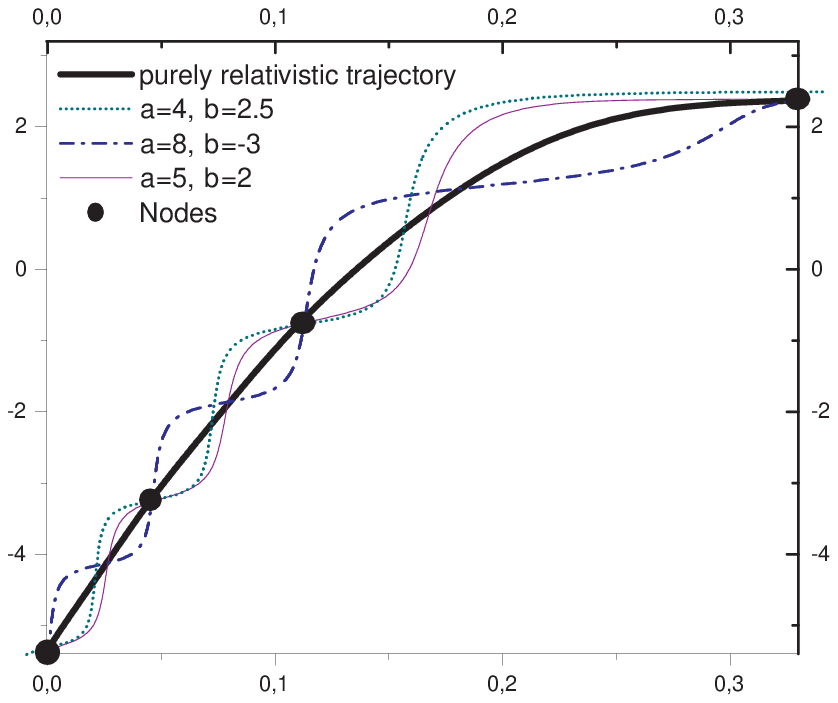}}%
\put(8.5,0.6) {$t$ ($\times 10^{-20}$ s)}%
\put(0.2,8.0) {$x$ ($\times 10^{-12}$ m)}%
}}
%\medskip
%\scriptstyle
\centerline { \vbox { \hsize=12cm\noindent \small Fig. 4:
Relativistic quantum trajectories for an electron of energy $E=2$
Mev moving under a constant potential. We have chosen $x(t=0)=
-5.4 \times 10^{-12}$ m.} }
\end{figure}

We would like to stress that as for the linear potential in
quantum cases, we check that the positions of the nodes on $x$
axis is related to the zeros  of the solution of the
Schr\"odinger equation $\phi_2$ present in the denominator of the
rapport in the expression (2) of the reduced action. This fact
indicates that, the RQTs likes the QTs and it is obvious that the
QT are a limit of the RQTs when $c \to \infty$.

% %%%%%%%%%%%%%%%%%%%%%%%%%%%%%%%%%%%%%%%%%%%%%%%%%%%%%%%%%%%%%%%%%
\vskip0.5\baselineskip
\noindent
{\bf 5.\ \ De Broglie's wavelength }
\vskip0.5\baselineskip
% %%%%%%%%%%%%%%%%%%%%%%%%%%%%%%%%%%%%%%%%%%%%%%%%%%%%%%%%%%%%%%%%%

The most important idea that Secs. 2, 3 and 4 bring
is the existence of node through which all RQTs pass, even
the purely relativistic one. In this section, we link
the distance between two adjacent nodes to the de Broglie's
wavelength
\begin {equation}
\lambda = {h \over p}
\end {equation}
In Eq. (25), $p$ is the relativistic momentum. For
a particle moving under a constant potential
\begin {equation}
p = {\sqrt{(E-U_0)^2-m_0^2c^4} \over c} \; .
\end {equation}
By replacing Eq. (26) in Eq. (25), we get
\begin {equation}
\lambda = {hc \over \sqrt{(E-U_0)^2-m_0^2c^4}} \; .
\end {equation}
The distance between two in the case of a constant
potential is
\begin {equation}
\Delta x_n={\pi \hbar c  \over
\sqrt{(E-U_0)^2-m_0^2c^4}}\; .
\end {equation}
From Eqs. (27) and (28) we get
\begin {equation}
\Delta x_n= {\lambda \over 2} \; .
\end {equation}
Thus, the de Broglie's wavelength represent the double of the
distance between two adjacent nodes. As we have presented in Ref.
\cite{Dja2}, we can generalize this definition for other
potentials. Indeed, if we compute the mean value of $\partial S_0
/ \partial x$ between two adjacent nodes, and taking into account
Eqs (11), (13) and (14) we find
\begin {equation}
\left<{\partial S_0 \over \partial x }\right> \equiv
{1 \over \Delta x_n} \int_{x(t_n)}^{x(t_{n+1})}
{\partial S_0 \over \partial x } \; dx
= {S_0(x(t_{n+1}))-S_0(x(t_n)) \over \Delta x_n} =
{\sqrt{(E-U_0)^2-m_0^2c^4} \over c} \; ,
\end {equation}
which is equal to $p$ (Eq. (26)). We propose to define a
new wavelength after substituting $p$ by
\begin {equation}
p = \left<{\partial S_0 \over \partial x }\right> \; .
\end {equation}
Then for any potential we can write, after using (2)
\begin {equation}
p = {\pi \hbar \over \Delta x} \; ,
\end {equation}
with $\Delta x$ is the distance between two adjacent nodes.
If we substitute (32) in (25) we find
\begin {equation}
\Delta x= {\lambda \over 2} \; .
\end {equation}
This relation links between the distance separating
two adjacent nodes and the de Broglie's wavelength.

% %%%%%%%%%%%%%%%%%%%%%%%%%%%%%%%%%%%%%%%%%%%%%%%%%%%%%%%%%%%%%%%%%
\vskip\baselineskip \noindent {\bf Conclusion} \vskip\baselineskip
% %%%%%%%%%%%%%%%%%%%%%%%%%%%%%%%%%%%%%%%%%%%%%%%%%%%%%%%%%%%%%%%%%

To conclude, we note that this article is a consequence of Ref.
\cite{Dja3}. We have showed that for the RQTs the nodes still as
an important result of our deterministic approach of quantum
mechanics. These nodes are linked successfully to the de
Broglie's wavelength. We also approach the photon's trajectories
for a constant potential. For photons, other potentials will be
studied in a next preparing paper.

By this results, we think that the generalization of our
formalism \cite{Dja1,Dja2,Dja3} into relativistic systems is
successful. Nevertheless, a generalization into more than one
dimension of our new approach of quantum mechanics is an
important step that we must investigate. In Ref. \cite{Dja4}, we
have started such as generalization by studying the QSHJE in three
dimensions for symmetrical potentials.

\newpage

% %%%%%%%%%%%%%%%%%%%%%%%%%%%%%%%%%%%%%%%%%%%%%%%%%%%%%%%%%%%%%%%%%
\vskip\baselineskip
\noindent
{\bf REFERENCES}
%\vskip\baselineskip
% %%%%%%%%%%%%%%%%%%%%%%%%%%%%%%%%%%%%%%%%%%%%%%%%%%%%%%%%%%%%%%%%%

\end {document}